\documentclass[useAMS,usenatbib,fleqn]{mn2e}

\usepackage{graphicx}
\usepackage{amsmath,amsfonts,amssymb}
\usepackage{color}
\usepackage{epsfig}

\usepackage{hhline}
\usepackage{array,booktabs}
\usepackage{epstopdf}
\usepackage[T1]{fontenc}

\newcommand{\msun}{M$_{\rm \odot}$}
\newcommand{\avgm}{$\langle m_{\rm *}\rangle$}
\newcommand{\apj}{ApJ}
\newcommand{\mnras}{MNRAS}
\newcommand{\aap}{A\&A}
\newcommand{\aj}{AJ}

\title[Compact Binary Mergers]{Compact Binaries Ejected from Globular Clusters as GW Sources}
\author[Bae, Kim \& Lee]{Yeong-Bok Bae$^{1}$\thanks{Email: baeyb@astro.snu.ac.kr}, Chunglee Kim$^{1}$\thanks{Email: chunglee@astro.snu.ac.kr} and Hyung Mok Lee$^{1,2}$\thanks{Email: hmlee@astro.snu.ac.kr}\\
$^{1}$Astronomy Program, Department of Physics and Astronomy, Seoul National University, 1 Gwanak-ro, Gwanak-gu, Seoul 151-742, Korea\\
$^{2}$Centre for Theoretical Physics, Seoul National University,  1 Gwanak-ro, Gwanak-gu, Seoul 151-742, Korea}

\begin{document}

\date{Accepted 2014 February 24.  Received 2014 February 24; in original form 2013 August 6}

\pagerange{\pageref{firstpage}--\pageref{lastpage}} \pubyear{2014}
\maketitle

\label{firstpage}

\begin{abstract}
Performing N-body simulations, we examine the dynamics of BH-BH (10 \msun~each) and NS-NS (1.4 \msun~each) binaries formed in a cluster and its implications for gravitational wave detection. A significant fraction of compact binaries are ejected from a globular cluster after core collapse. Among the total number of ejected compact objects, 30 per cent of them are in binaries. Merging time-scales of ejected binaries, which depend on the cluster's velocity dispersion, are in some cases shorter than the age of the universe. During the merging event, these dynamically formed compact mergers are expected to produce gravitational waves that can be detectable by the advanced ground-based interferometers. Based on our reference assumptions, merger rates of ejected BH-BH and NS-NS binaries per globular cluster are estimated to be 2.5 and 0.27 per Gyr, respectively. Assuming the spatial density of globular clusters to be 8.4$h^3$ clusters Mpc$^{-3}$ and extrapolating the merger rate estimates to the horizon distance of the advanced LIGO-Virgo network, we expect the detection rates for BH-BH and NS-NS binaries with cluster origin are to be 15 and 0.024 yr$^{-1}$, respectively. We find out that some of the dynamically formed binaries are ejected with a large escape velocity. They can be responsible for short gamma-ray bursts whose locations are far from host galaxies. 
\end{abstract}


\section{Introduction}

Compact binary coalescences (CBCs) are important sources of gravitational waves (GWs) as well as electromagnetic bursts (e.g., supernovae and gamma-ray bursts). The formation mechanism of compact binaries is largely classified into two categories and one of them is from isolated binary evolution. The empirical method of extrapolating the pulsar measurements is used for estimating NS-NS merger rates \citep*{kim03, kim10, kalogera2004} and population synthesis method is also used to constrain the number of black holes (BHs) as well as neutron stars (NSs) and evolution of binaries in field populations \citep{bel07, dominik2012}. It is also known that compact binaries can be formed through the dynamical interactions with surrounding stars in globular clusters (GCs) (e.g., \citealt{sig93,lee01}). Self-gravitating systems such as GCs undergo core collapse and time-scale of that decreases if massive components exist \citep{spi87}. Due to the short life span of massive main-sequence stars, BHs become relatively high mass components among the stars and they fall into the core region through the dynamical frictions. In high density core region, BH-BH binaries are formed through three-body process or exchanging the pair of existing binaries. The interactions of BH binaries and surrounding components make the binary's orbit more compact \citep{heg75,good93} and provide significant kick velocities. Finally, ejection occurs when their recoil velocity becomes greater than the escape velocity of the GC. These ejected binaries are generally tightly bounded and some of them will merge by GW emission within the age of the universe (e.g., \citealt{dow10,dow11,tan13}).

It is expected that NS-NS merger rate predominates over BH-BH or BH-NS merger rate in the Galactic field environment (e.g., \citealt{bel07}). The estimated detection rate of NS-NS merger through advanced LIGO-Virgo networks is also larger than that of BH-BH or BH-NS merger in the field \citep{aba10}, despite the smallest detection volume of NS-NS merger. In contrast, the merger rate of dynamically formed BH-BH binaries is estimated to exceed NS-NS or BH-NS merger rate from GCs. The binaries containing NSs can hardly contribute to merger rate originated from GCs, because more massive BHs substitute for NSs in binary during the interactions. The merger rate of dynamically formed NS-NS binaries is estimated much lower than that from the field \citep{phinney91}. Therefore, most of the previous studies paid attention to only BH-BH binaries when they consider the mergers from GCs. 

In this work, we examine binary mergers that are dynamically formed in and ejected from GCs, including not only BH-BH binaries but also BH-NS and NS-NS binaries. In \S \ref{sec:model}, we describe our simplified cluster model and assumptions used in N-body simulations. In \S \ref{sec:results}, we present results focusing on the properties of `ejected' compact binaries (BH-BH and NS-NS). In \S \ref{sec:rates}, we present a table of merger rate estimate and the expected detection of GWs associated these binaries for the network of Earth-bound interferometers. In \S \ref{sec:discussions}, we further discuss our results and their implications for a subset of short GRBs that shows large offsets from host galaxies.

\begin{figure} 
\centering
\includegraphics[width=1\columnwidth]{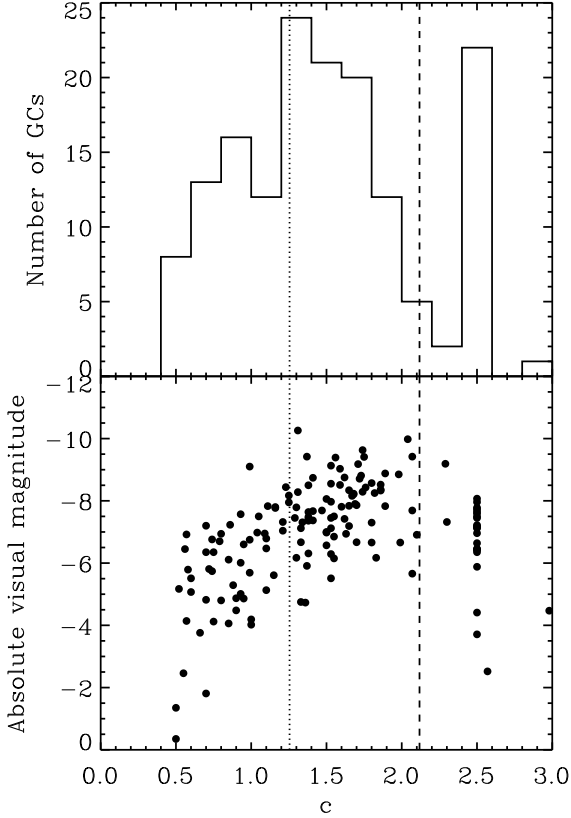}
\caption{Histogram of concentrations (the upper panel) and distribution of absolute visual magnitudes (the lower panel) of GCs in the Milky Way listed in the Harris' catalogue \citep{har10}. The concentration ($c$) is essentially a ratio between the tidal radius ($r_{\rm t}$) and core radius ($r_{\rm c}$) of the cluster ($c\equiv {\rm log}_{10}(r_{\rm t}/r_{\rm c})$). Note $c=2.5$ is an arbitrarily given value in Harris' catalogue for some of the core collapsed clusters. Dotted and dashed lines in both panels represent $W_{0}=6$ and $W_{0}=9$, respectively.}
\label{fig:gc_concen}
\end{figure}

\begin{table}  
\caption{Models used in this work. The first column shows model names (see text for description of labelling). The second column is the $W_{0}$ value of the King model. The next three columns are the total number of particles ($N_{\rm tot}$) and the number of NSs and BHs ($N_{\rm NS}$, $N_{\rm BH}$), respectively. $f_{\rm pb}$ represents the number fraction of stars in primordial binaries and $N_{\rm run}$ is the number of simulations we performed per model.}
\label{table:models}
\begin{tabular}{@{}lcccccc}
\toprule
Model & $W_{0}$ & $N_{\rm tot} (\times 1000) $ & $N_{\rm NS}$ & $N_{\rm BH}$ & $f_{\rm pb}$ & $N_{\rm run}$ \\
\toprule
A5kN5      & 6       & 5       & 128      & 0        & 0       & 8         \\[1.5ex]

A10kN1     & 6       & 10       & 50       & 0        & 0       & 10        \\
A10kN2     & 6       & 10       & 101      & 0        & 0       & 8         \\
A10kN5     & 6       & 10       & 256      & 0        & 0       & 6         \\
A10kN10    & 6       & 10       & 526      & 0        & 0       & 4         \\[1.5ex]

A10kN5p10  & 6       & 10       & 256      & 0        & 0.1     & 4         \\
A10kN5p30  & 6       & 10      & 256      & 0        & 0.3     & 4         \\
A10kN5p50  & 6       & 10      & 256      & 0        & 0.5     & 4         \\[1.5ex]

A20kN5     & 6       & 20   & 513      & 0        & 0       & 2         \\[1.5ex]

A25kB5     & 6       & 25     & 0        & 92       & 0       & 6         \\[1.5ex]

A25kB5p10  & 6       & 25       & 0        & 92       & 0.1     & 3         \\
A25kB5p30  & 6       & 25       & 0        & 92       & 0.3     & 3         \\
A25kB5p50  & 6       & 25       & 0        & 92       & 0.5     & 3         \\[1.5ex]

A50kB2     & 6       & 50     & 0        & 71       & 0       & 6         \\
A50kB5     & 6       & 50      & 0        & 184      & 0       & 3         \\[1.5ex]

\midrule
A50kBN     & 6       & 50      & 250      & 100      & 0       & 1         \\
\midrule
B5kN5      & 9       & 5    & 128      & 0        & 0       & 8         \\[1.5ex]

B10kN1     & 9       & 10      & 50       & 0        & 0       & 10        \\
B10kN2     & 9       & 10      & 101      & 0        & 0       & 8         \\
B10kN5     & 9       & 10      & 256      & 0        & 0       & 6         \\
B10kN10    & 9       & 10       & 526      & 0        & 0       & 4         \\[1.5ex]

B10kN5p10  & 9       & 10       & 256      & 0        & 0.1     & 4         \\
B10kN5p30  & 9       & 10       & 256      & 0        & 0.3     & 4         \\
B10kN5p50  & 9       & 10       & 256      & 0        & 0.5     & 4         \\[1.5ex]

B20kN5     & 9       & 20    & 513      & 0        & 0       & 2         \\[1.5ex]

B25kB5     & 9       & 25       & 0        & 92       & 0       & 6         \\[1.5ex]

B25kB5p10  & 9       & 25       & 0        & 92       & 0.1     & 3         \\
B25kB5p30  & 9       & 25       & 0        & 92       & 0.3     & 3         \\
B25kB5p50  & 9       & 25       & 0        & 92       & 0.5     & 3         \\[1.5ex]

B50kB2     & 9       & 50     & 0        & 71       & 0       & 6         \\
B50kB5     & 9       & 50     & 0        & 184      & 0       & 3         \\[1.5ex]
\bottomrule
\end{tabular}
\end{table}

\section{Model}\label{sec:model}

Modelling of cluster's evolution involves external tidal field, mass function, primordial binaries, stellar evolution, and so on. There are limitations, however, in performing N-body simulations in order to realize a GC. In particular, large number of stars of actual GC ($\sim10^{6}$) can hardly be achievable. Therefore, our model is simplified with various assumptions and careful selections of model parameters.

We use King model \citep{king66} with a static external tidal field as a GC's initial density profile. The external tidal field is created from a point mass galaxy whose circular velocity is 220 km s$^{-1}$ at a distance of 8.5 kpc from the centre.

The $W_{0}$ parameter of the King model, defined by the ratio between a central potential and velocity dispersion, represents the concentration level of a GC. The more concentrated cluster has a larger value of $W_{0}$. The upper panel of Fig.~\ref{fig:gc_concen} shows a histogram of estimated concentrations ($c$) of known GCs in the Milky Way (MW). Data are taken from the Harris' catalogue \citep{har10}. The mean concentration of MW GCs, except those with arbitrary value of $c=2.5$ in the catalogue, is about 1.34. This is slightly larger than $c=1.25$ that corresponds to our reference value of $W_{0}=6$ (dotted line). The lower panel of Fig.~\ref{fig:gc_concen} represents absolute visual magnitudes of the clusters. The massive (i.e., brighter; recall mass-to-light ratio \citep{McL05}) GCs are more likely to be strongly concentrated. Massive GCs are expected to produce more compact binaries than lighter ones. Therefore, we consider more concentrated models with $W_{0}=9$ in addition to those with $W_{0}=6$.
 
For a cluster that is around several tens of Myrs old, almost all massive stars in GCs have finished their evolution and turned to compact remnants such as BHs and NSs after the last supernova explosion. At this stage, the cluster also contains dwarf stars and main-sequence stars whose mass is smaller than the turnoff mass of the GC. A turnoff mass estimated from 17 MW GCs, based on the multi-mass King model, is about 0.8 \msun~\citep{pau10}. In this work, we adopt a simplified initial mass function (IMF), introducing only three mass components: 10 \msun~BHs, 1.4 \msun~NSs, and 0.7 \msun~main-sequence stars. If we assume the masses of NS and BH progenitors to be $8-20$ \msun~and $\ge 20$ \msun, respectively and applying the Kroupa IMF \citep{kro01} between 0.08 and 150 \msun~\citep{wei04}, the number fractions of NSs and BHs are about 0.45 and 0.18 per cent of the total number of stars in a cluster. In this case, the mass fractions of NSs ($f_{\rm NS}$) and BHs ($f_{\rm BH}$) are about 1 and 2.5 per cent of the total cluster mass if a cluster has only three mass components as stated earlier. 

Lastly, clusters containing primordial binaries are also studied. The fraction of stars in primordial binaries ($f_{\rm pb}$) is assumed to be 10, 30, or 50 per cent. For example, a model with $f_{\rm pb}=0.5$ contains $0.5\times N_{\rm tot}/2$ primordial binaries. Eccentricities of primordial binaries are assumed to follow thermal distribution $f(e)\sim2e$. The semi-major axes are determined by Kroupa period distribution \citep{kro95}
\begin{equation}
f(P)=2.5\frac{{\rm log}_{10}P-1}{45+({\rm log}_{10}P-1)^2}~,
\end{equation}
where $P$ is an orbital period in days. ${\rm log}_{10}P_{\rm min}=1$ and ${\rm log}_{10}P_{\rm max}=8.43$ are adopted as minimum and maximum periods. A pair of primordial binary is established by random selection of two stars in a cluster regardless of their masses.

We use \textsc{nbody}6 code \citep{aar03,nitadori12} that enables us for using graphical processing units (GPUs) on NVIDIA Tesla c1060 platforms. We perform simulations for model clusters, varying total number of stars ($N_{\rm tot}$) and mass fractions of BHs or NSs (1, 2, 5, and 10 per cent). We establish 5k$-$50k stars in a cluster, where `k' stands for a thousand. For clusters containing BHs, we model $N_{\rm tot}=25$k or 50k in order to ensure the sufficient number of BHs to interact. In addition, we consider models with exaggerated mass fractions of BHs or NSs (5 and 10 per cent) considering uncertainties in mass fraction as well as the limitation in $N_{\rm tot}$ in N-body simulations. Each of three mass components (10 \msun~BHs, 1.4 \msun~NSs or 0.7 \msun~main-sequence stars) is initially set to have density distribution of King model.

Cluster models used in this work are listed in Table~\ref{table:models}. A model name starts with a capital letter `A' ($W_{0}=6$) or `B' ($W_{0}=9$) representing the level of concentration. The total number of particles used in each model ($N_{\rm tot}$) are indicated as 5k, 10k, 20k, 25k and 50k. The capital `N' or `B' with numbers (1, 2, 5, or 10) means the mass fraction of BHs or NSs in per cent. The number of BHs or NSs for each model shown in the Table 1 ($N_{\rm BH}$ or $N_{\rm NS}$) is obtained with a fixed mass fraction. Except the model A50kBN where all three stellar populations exist in a cluster, all models are composed of stars with either BHs or NSs. The lower-case letter `p' with numeric values of 10, 30, or 50 represents the fraction of primordial binaries in the model cluster.

It is known that many radio pulsars have large peculiar velocities attributed to the asymmetric supernova explosions \citep*{wjm2013}. The empirical kick velocity distribution obtained from the proper motion measurements of young radio pulsars with ages less than 3 Myr has a peak around a few hundred km s$^{-1}$ \citep{hobbs2005}. There are also a number of pulsars with transverse velocities of a thousand km s$^{-1}$, most notably the hyperfast PSR B1508+55 \citep{cha05}. \citet{lyn94} and \citet*{lor97} showed that the mean birth velocity of pulsars is $450-500$ km s$^{-1}$. \citet{han97} found that the velocity distribution could be represented by a Maxwellian with velocity dispersion of 190 km s$^{-1}$ and mean value of $250-300$ km s$^{-1}$. \citet{har97}, however, found that there are also many low velocity pulsars as well as high velocity pulsars. Moreover, significant fraction of pulsars formed in GCs are likely to be associated with either symmetric or `small-kick' explosions (\citealt{freire2013} and references therein). \citet*{fry98} argued for a bimodal distribution, roughly 30 per cent of NSs receive no kick but the rest 70 per cent of NSs receive a large kick over 600 km s$^{-1}$. \citet*{arz02} also suggested two Gaussian components, 90 km s$^{-1}$ and 500 km s$^{-1}$ for 40 and 60 per cent of NSs, respectively. 

Meanwhile, the magnitude of BH natal kick is highly uncertain \citep*{nelemans99, jonker04, willems05, dhawan07, fragos09, rep12, wong12, wong13}. \citet{rep12} found that natal kick velocities of BHs are similar to those of NSs using the distribution of low-mass X-ray binaries containing BHs within our Galaxy. \citet{willems05} suggested the upper limit of BH kick as $\approx210$ km s$^{-1}$ for GRO J1655-40 and \citet{fragos09} derived $80-300$ km s$^{-1}$ of BH natal kick for XTE J1118+480. On the other hand, \citet{nelemans99} and \citet{dhawan07} found that the natal kick of BH does not need to explain the space velocities of BH X-ray binaries. In addition, \citet{wong12} suggested relatively low natal kick velocity for the case of Cygnus X-1 ($\leq77$km s$^{-1}$ at 95 per cent confidence). In this work, we do not apply the natal kicks directly to the simulations both for NSs and BHs. We assume that all NSs and BHs are retained in a cluster after supernova explosions and follow the velocity dispersion of King model. The case including natal kicks will be discussed in \S \ref{sec:rates}.

Note that stellar evolution is not included in this work. We focus on the dynamical evolution of model clusters, starting from the time all massive stars in a cluster have completed their evolution. The evolution time-scale of normal stars (with a fixed mass of 0.7 \msun) is much longer than that of massive stars. Hence, low-mass stars' evolution can be neglected for a cluster right after the last supernova explosion that sets $t=0$ in our modelling.

\section{Results}\label{sec:results}

\subsection{Ejection of compact binaries}

\begin{figure} 
\centering
\includegraphics[width=1\columnwidth]{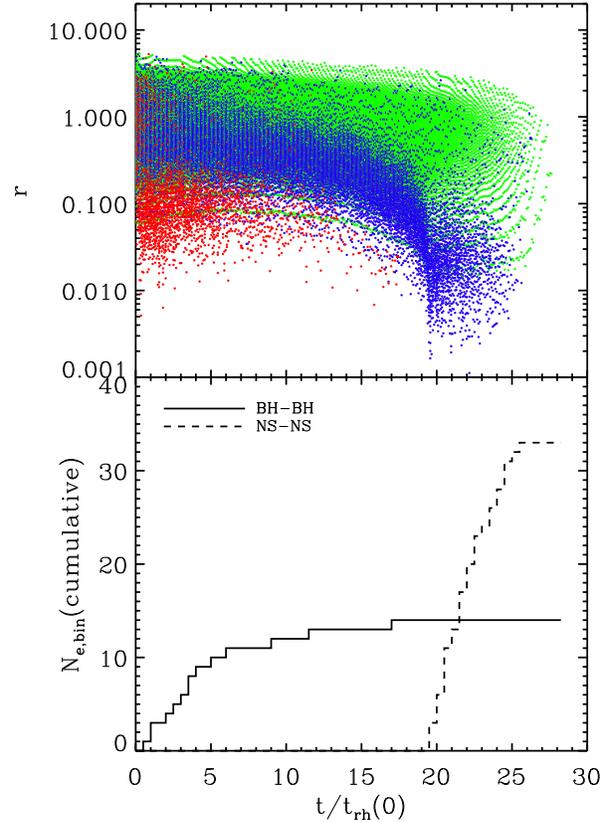}
\caption{The upper panel: The distance of each mass component from the cluster centre is plotted against time. The total number of particles in the model is 50k, including 250 NSs and 100 BHs (Model: A50kBN). Green, blue, and red dots stand for the mean distance obtained from adjacent 200 ordinary stars, 4 NSs, and 2 BHs, respectively. The lower panel: The cumulative number of ejected BH-BH or NS-NS binaries over time obtained from the same model.}
\label{fig:posi_nesc}
\end{figure}

Fig.~\ref{fig:posi_nesc} shows the dynamical evolution of a cluster model A50kBN. This is the only model containing all three mass components. The number of NSs and BHs in this model cluster are 250 and 100, respectively. The NS and BH number fractions are similar to what is expected from the Kroupa IMF \citep{kro01}. The time $t$ in the figure is normalized by the cluster's initial half-mass relaxation time $t_{\rm rh}$ \citep{spi87}
\begin{equation} 
t_{\rm rh} = 0.138 \frac{N^{1/2} r_{\rm h}^{3/2}}{G^{1/2} \langle m_{\star}\rangle^{1/2} ln(\gamma N)}~,
\end{equation}
where $N$ and $r_{\rm h}$ are the total number of stars and a half-mass radius of the cluster, respectively and $\langle m_{\rm *}\rangle$ represents the average mass of the stars in the cluster. We use $\gamma=0.11$ for the Coulomb logarithmic value \citep{gie94}.

The upper panel shows distances of BHs (red dots), NSs (blue dots), and the rest 0.7 \msun~stars (green dots) from the cluster centre and it depicts the dynamical evolution of each mass components. First of all, BHs sink toward the central region of a cluster very quickly because they are the most massive components. The cluster reaches a quasi-stable state against core collapse until a significant fraction of BHs are ejected, either in binaries or as singles, through binary-single interactions. The migration of NSs to the centre is delayed by heating from BH-BH binaries in the central region during this phase. Core collapse driven by NSs resumes after BHs are depleted; NSs sink into the centre, form binaries, and get ejected by binary-single interactions. The lower panel shows the number of ejected binaries from the cluster ($N_{\rm e,bin}$). The solid line shows BH-BH binaries and dashed line represents NS-NS binaries. The formation and ejection of compact binaries start with core collapse. We emphasize that NSs start to sink only when almost all BHs are ejected from the cluster. This implies that the interactions between BHs and NSs are rare. We find no BH-NS ejected binary formed in the model A50kBN. As we confirm the interaction time of BHs and NSs is well separated from each other from this exercise, we include only BHs or NSs to cluster models discussed below.

\begin{figure*} 
\centering
\includegraphics[width=0.9\textwidth]{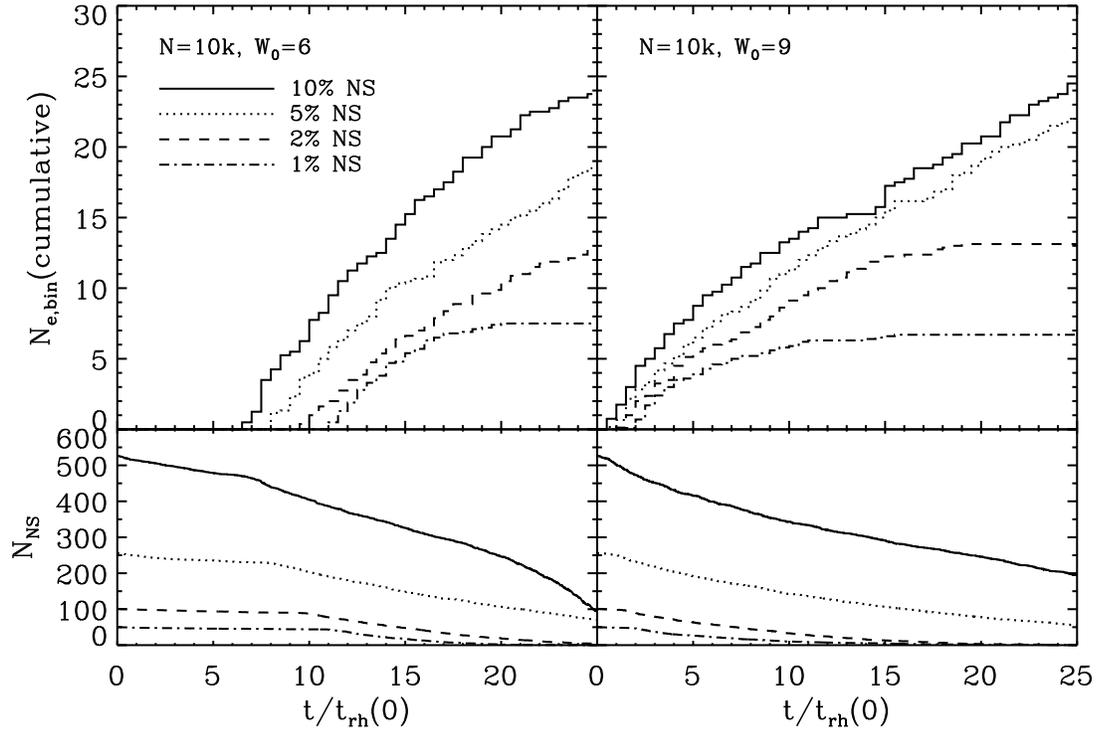}
\caption{The upper panel: The cumulative number of ejected NS-NS binaries obtained from clusters with different NS mass fractions. The lower panel: Evolution of the number of NSs in a tidal radius.}
\label{fig:nesc_nn}
\end{figure*}

Fig.~\ref{fig:nesc_nn} shows results obtained from clusters with different NS mass fractions, i.e., 1, 2, 5, and 10 per cent where the total number of particles is fixed to be 10k. The upper panels (left and right) present an evolution of the cumulative number of ejected binaries for clusters with $W_{0}=6$ and 9. The results show that the NS mass fraction affects a cluster's core collapse time and hence, governs the time when a compact binary is first ejected from a cluster. As shown in the figure, core collapses of clusters with $W_{0}=6$ normally occur around $t=8$ $t_{\rm rh}(0)$. More `concentrated' clusters with $W_{0}=9$ go through core collapses much quickly ($t\approx1$ $t_{\rm rh}(0)$) than those with $W_{0}=6$.

\begin{figure*} 
\centering
\includegraphics[width=0.9\textwidth]{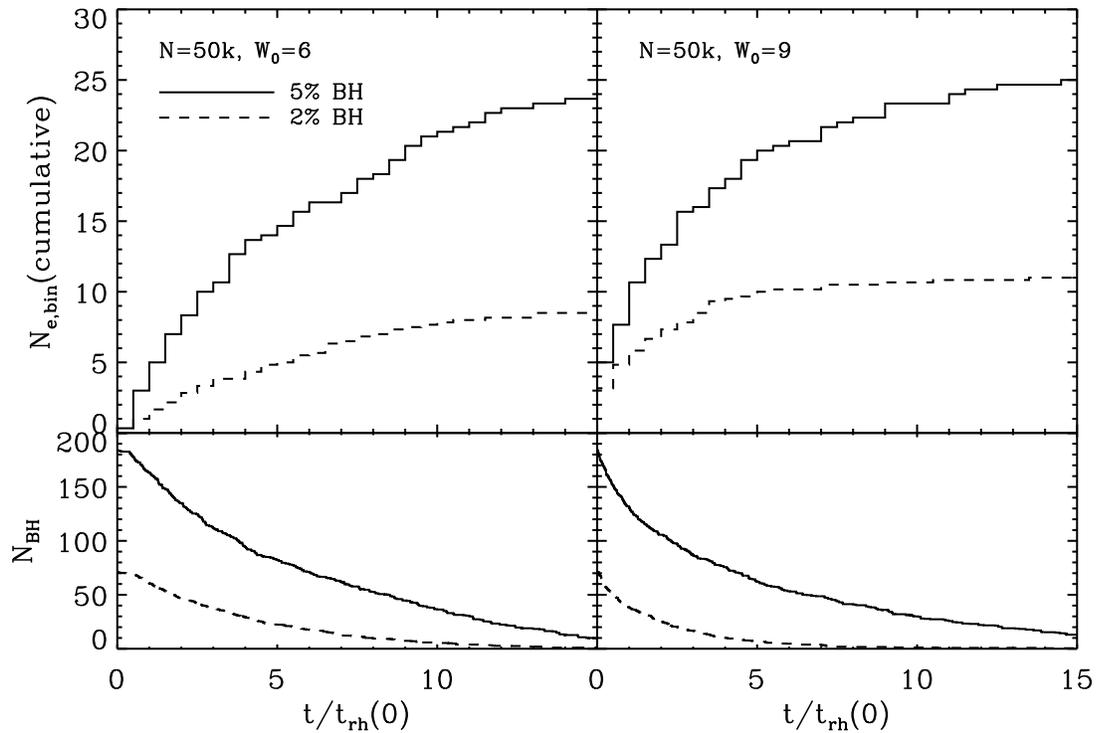}
\caption{Same as Fig.~\ref{fig:nesc_nn}, but for BHs.}
\label{fig:nesc_bb}
\end{figure*}

\begin{figure*} 
\centering
\includegraphics[width=1.05\textwidth]{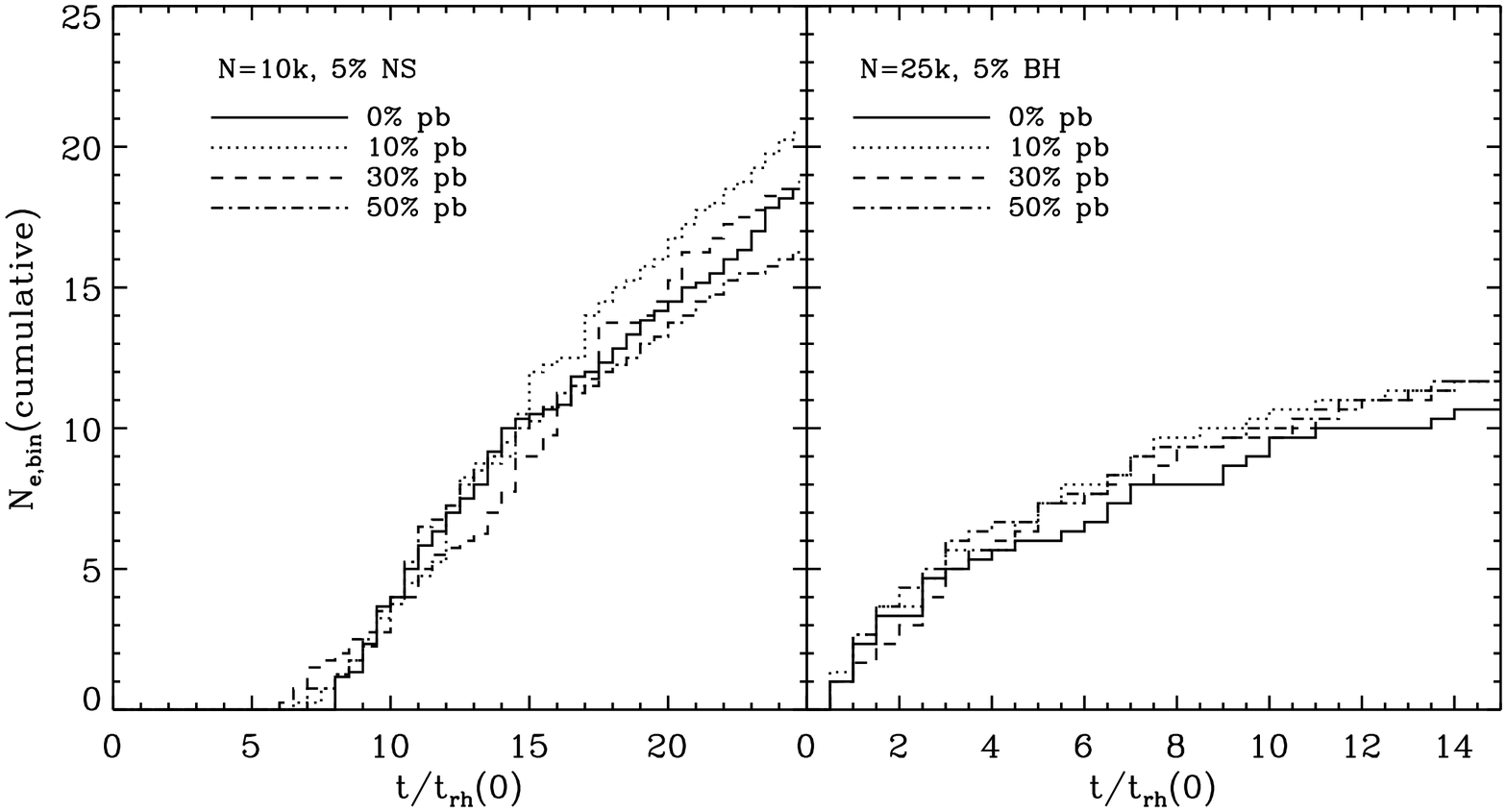}
\caption{The number of ejected NS-NS (left panel) or BH-BH binaries (right panel) obtained from cluster models with different fractions of primordial binaries (pb). Each model has a fixed concentration of $W_{0}=6$ and a mass fraction of 5 per cent for the compact objects. A 1-dimensional central velocity dispersion is also fixed to be 3 km s$^{-1}$ (left panel) or 5 km s$^{-1}$ (right panel), respectively. The number of ejected binaries from a cluster is barely affected by the fraction of primordial binaries.}
\label{fig:nesc_pb}
\end{figure*}

The ejection rates of NS-NS binaries can be obtained from slopes of the cumulative number of ejected NS-NS binaries (the upper panels of Fig.~\ref{fig:nesc_nn}). We find that the ejection rates of NS-NS binaries are almost the same in early stages of clusters regardless of the assumption on the NS fraction. This is because the number of NSs to interact in the central region is confined, even though the fraction of NSs increases. However, in late stages of a cluster, ejection rates of NS-NS binaries decreases for clusters with lower NS fractions due to depletion of NSs (models: A10kN1, B10kN1, B10kN2). 

The lower panels of Fig.~\ref{fig:nesc_nn} represent the number of NSs located in a tidal radius. By comparing upper and lower panels, we obtain ratios between the number of ejected NSs in NS-NS binaries and the total number of ejected NSs ($f_{\rm e,NN}$). For clusters with $W_{0}=6$, we obtain $f_{\rm e,NN}=11, 20, 26$, and 30 per cent at $t=25$ $t_{\rm rh}(0)$ by varying $f_{\rm NS}=10, 5, 2$, and 1 per cent, respectively. For more concentrated clusters ($W_{0}=9$), we obtain $f_{\rm e,NN}=15, 22, 26$, and 27 per cent from models with $f_{\rm NS}=10, 5, 2$, and 1 per cent, respectively. As $f_{\rm NS}$ decreases to the predicted value of $\sim 1$ per cent based on the Kroupa IMF, $f_{\rm e,NN}$ asymptotically approaches to 30 per cent regardless of the assumptions on $W_{0}$.

We show results obtained from clusters containing BHs in Fig.~\ref{fig:nesc_bb}. In general, the results are similar to what we described earlier for models with NSs. One of the main differences between clusters with NSs and BHs is the onset of core collapse, i.e., the formation and ejection of binaries. Core collapse of a cluster occurs faster, if the mass ratio between the heavy mass component (BH or NS) and light one (0.7 \msun~star) is large. Based on our model parameters, BHs are much heavier than NSs (10 \msun~vs 1.4 \msun). Thus, core collapses of clusters containing BHs occur faster than those containing NSs, before $t=1$ $t_{\rm rh}(0)$ regardless of $W_{0}$. In addition, unlike what we observe from Fig.~\ref{fig:nesc_nn}, the number of ejected BH-BH binaries as well as their ejection rates are proportional to the BH fraction in a cluster. This is because the number of BHs interacting in the central region of a cluster increases as the fraction of BHs increases. Interestingly, the ratios between the number of ejected BHs in BH-BH binaries and the total number of ejected BHs ($f_{\rm e,BB}$) are about 30 per cent similar to what we find from  Fig.~\ref{fig:nesc_nn}. $f_{\rm e,BB}$ obtained from clusters with $f_{\rm BH}=5$ and 2 per cent at $t=15$ $t_{\rm rh}(0)$ are 27, 24 per cent ($W_{0}=6$) and 29, 31 per cent ($W_{0}=9$), respectively. The rest of BHs are ejected in singles or in binaries pairing with 0.7 \msun~stars. 

As we explained earlier in the previous section, we also vary the total number of stars, fixing NS or BH fraction. The number of ejected NS-NS or BH-BH binaries is proportional to the total number of stars in a cluster. The ratio of ejected BHs or NSs in the form of binary is also about 30 per cent regardless of the total number of stars. Therefore, we extrapolate these results to MW GCs that consist of hundreds of thousands of stars when we consider merger rates of ejected compact binaries in \S \ref{sec:rates}.

Lastly, we examine the effects of primordial binaries on the number of ejected compact binaries from a cluster. The presciption to establish primordial binaries in a cluster is described in the previous section. We fix the NS or BH mass fraction to be 5 per cent. The 1-dimensional central velocity dispersion of a cluster is fixed to be 3 km s$^{-1}$ (for clusters containing NSs) or 5 km s$^{-1}$ (for those containing BHs). Then, we vary the number fraction of primordial binaries (0, 10, 30 and 50 per cent). The results are given in Fig.~\ref{fig:nesc_pb}. The number of ejected binaries does not depend on the fraction of primordial binaries. Even though the primordial binaries exist, most of them are weakly bound relative to the kinetic energy of surrounding stars. Therefore, primordial binaries are easily destroyed by the interactions with other stars or binaries. On the other hand, the number of ejected NS-NS or BH-BH binaries in clusters with lower central velocity dispersions ($\sim 1$ km s$^{-1}$) is weakly dependent on the fraction of primordial binaries because their binding energies relative to the kinetic energy of surrounding stars become larger. However, we discuss in the following section (\S \ref{sec:rates}), contribution from clusters with lower central velocity dispersions to the merger rate of compact binary is negligible. Overall, we conclude that the existence of primordial binaries does not affect the merger rate of binaries ejected from clusters.

\subsection{Properties of ejected compact binaries}

In this section, we describe properties of the ejected binaries: eccentricities, hardnesses, and velocities normalized by the averaged escape velocity of a cluster. These properties do not rely on the fraction of compact objects or total number of stars, but weakly depend on the concentration of the cluster except for the eccentricities. Therefore, we present distributions of these properties of ejected binaries for clusters with $W_{0}=6$ and 9.

\begin{figure} 
\centering
\includegraphics[width=1\columnwidth]{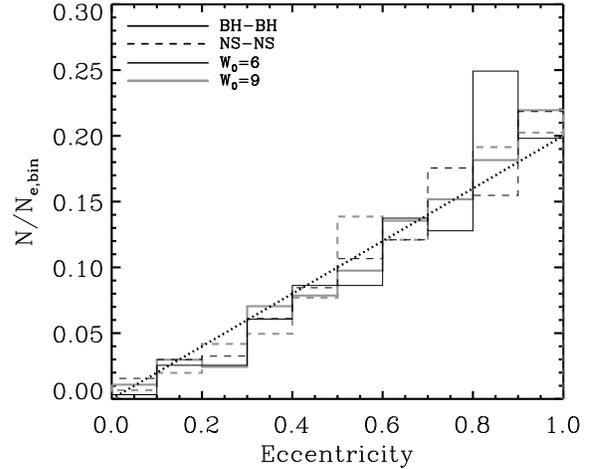}
\caption{The eccentricity distribution of ejected NS-NS (dashed) and BH-BH (solid) binaries from the clusters with $W_{0}=6$ (black) and $W_{0}=9$ (gray). They are explained well by a thermal distribution (dotted line).}
\label{fig:ecc}
\end{figure}

The eccentricity distribution of ejected binaries is invariant of cluster model parameters and is described as the so-called `thermal distribution', $f(e)\sim2e$ (Fig.~\ref{fig:ecc}). The majority of ejected binaries have large eccentricities, implying shorter merging time-scales (see \S \ref{sec:rates}). 

\begin{figure} 
\centering
\includegraphics[width=1\columnwidth]{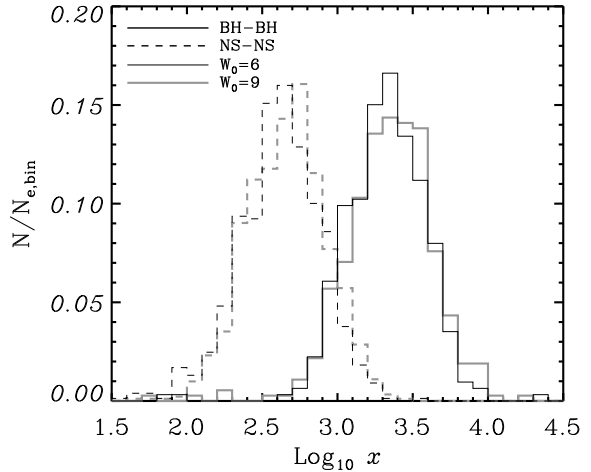}
\caption{The hardness distribution of ejected NS-NS (dashed) and BH-BH (solid) binaries from the clusters with $W_{0}=6$ (black) and $W_{0}=9$ (gray). Note the x-axis is in a logarithmic scale.} 
\label{fig:loghard}
\end{figure}

The dimensionless hardness of a binary is defined as follows
\begin{equation}  
\label{eq:hardness}
x=\frac{G m_{\rm 1} m_{\rm 2}/2a}{3 \langle m_{\rm *} \rangle \sigma^{2}/2}~,
\end{equation}
where $m_{1}$ and $m_{2}$ are masses of each stars and $a$ is a semi-major axis of the binary and $\langle m_{\rm *} \rangle$ is the average mass of all stars in the cluster. We calculate the velocity dispersion of a cluster $\sigma^{2}$, from the central region of a cluster which contains 25 per cent of total mass. The hardness of a binary represents its compactness or the binary's binding energy relative to the kinetic energy of surrounding stars. Fig.~\ref{fig:loghard} shows the hardness distribution of compact binaries ejected from clusters follows lognormal as expected \citep{ole06}. The average hardnesses of ejected NS-NS binaries are about $x=500$ ($W_{0}=6$) and 540 ($W_{0}=9$) for different cluster models. Similarily, hardnesses of BH-BH binaries are estimated to be $x \sim 2500$ ($W_{0}=6$) and $\sim2700$ ($W_{0}=9$), respectively. Hardnesses of binaries ejected from more concentrated clusters are about 10 per cent larger than those from less concentrated ones. Note that there are a few extremely hard binaries with $x > 2500$ (NS-NS) or $x \sim 25000$ (BH-BH). These hardest binaries are the best examples of merging binaries.

\begin{figure} 
\centering
\includegraphics[width=1\columnwidth]{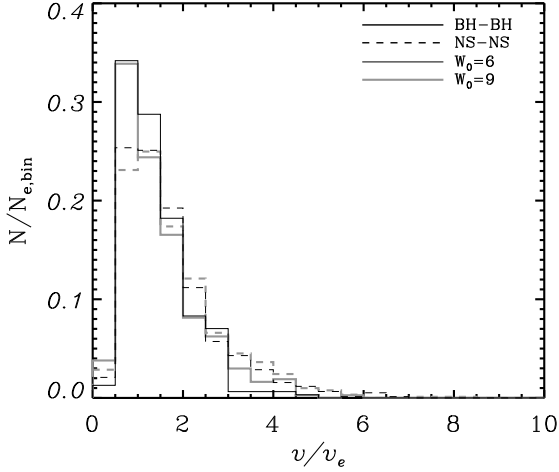}
\caption{The velocity distribution of ejected NS-NS (dashed) and BH-BH (solid) binaries from the clusters with $W_{0}=6$ (black) and $W_{0}=9$ (gray). The velocities are measured at the tidal radius and normalized by the averaged escape velocity ($v_{\rm e}$) of a cluster where the binaries are formed.}
\label{fig:vel_esc}
\end{figure}

The velocity distribution of ejected NS-NS or BH-BH binaries at the cluster's tidal radius when it gets out of a cluster is shown in Fig.~\ref{fig:vel_esc}. The velocity is normalized by the averaged escape velocity of the cluster, which is defined by $v_{\rm e}=\sqrt{\langle v_{\rm esc}^{2}\rangle}$, where
$\langle v_{\rm esc}^{2}\rangle=4v_{\rm m}^{2}$. Here, $v_{\rm m}$ is a root-mean-square three-dimensional velocity obtained from all types of stars in a cluster. The mean velocity normalized by $v_{\rm e}$ of NS-NS and BH-BH binaries are about 1.4 and 1.7 for the clusters with $W_{0}=6$ (1.5 and 1.8 for the clusters with $W_{0}=9$), respectively. We note that some of the ejected binaries have more than 5 times of the escape velocity, i.e., $150\sim200$ km s$^{-1}$. These are expected to move significantly far from the host cluster over a few Gyr.

\section{Merger Rate of NS-NS and BH-BH binaries ejected from clusters}\label{sec:rates}

Compact binary mergers are prime targets for GW detections with the advanced LIGO-Virgo network \citep{aLIGO, aVirgo}. The GW detection rate of compact binary mergers originated from clusters can be obtained by following procedures. First, we calculate the ratio of merging binaries among those ejected from model clusters and extrapolate the merger rate of NS-NS and BH-BH binaries to MW GCs. Then, the merger rate per GC is derived and the GW detection rate is calculated by multiplying the detection volume of the advanced detector network and the spatial density of GCs, where the merging binaries are produced. In this work, we define `merging binaries' as those with merging time-scales less than $\sim13$ Gyr, the age of the universe.

The merging time-scale of a binary is estimated by the Peters' formula \citep{pet64}, 
based on the assumption that GW emission is the sole energy loss mechanism of the binary.
\begin{equation}
\label{eq:tmerge}
t_{mg}=\frac{5}{64}\frac{c^5 a^4}{G^3 m_{1}m_{2}(m_{1}+m_{2})f(e)}~,
\end{equation}
where $f(e)$ is defined as follows
\begin{equation}
f(e)=(1-e^2)^{-7/2}\left({1+\frac{73}{24}e^2+\frac{37}{96}e^4}\right)~.
\end{equation}
Here, $m_{1}$ and $m_{2}$ are mass components in a binary, $a$ is a semi-major axis and $e$ is an eccentricity. Eccentric binaries or those in tight orbit (i.e, small $a$) have shorter merging time-scales. As we described earlier, we assume NS-NS or BH-BH binaries with same masses, i.e., $m_{1}=m_{2}=1.4$ or 10 \msun.

\begin{figure} 
\centering
\includegraphics[width=1\columnwidth]{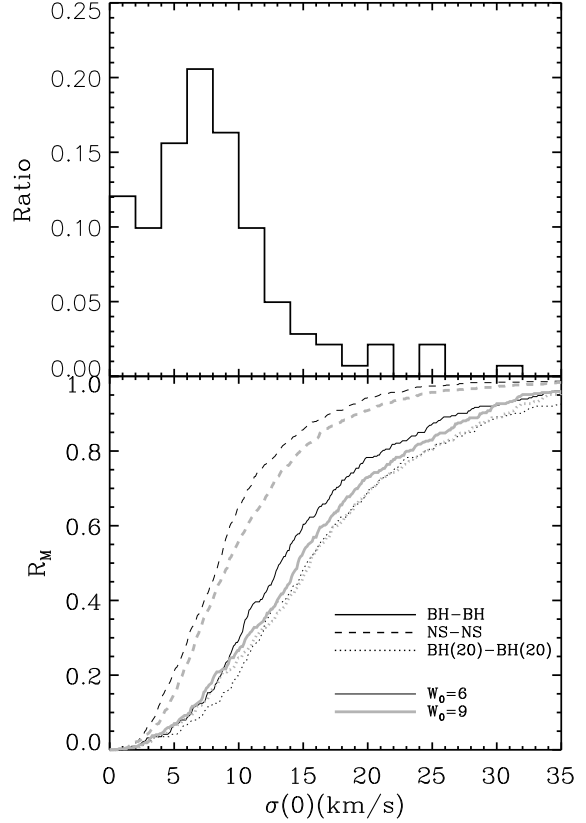}
\caption{The upper panel: a histogram of the central velocity of 141 GCs in MW \citep{gne02}. The lower panel: The ratio of merging binaries to the total ejected binaries as a function of the central velocity of a GC: NS-NS (black and gray dashed lines) and BH-BH (black and gray solid lines). Black ($W_{0}=6$) and gray ($W_{0}=9$) lines represent different cluster concentration. We also show the results from a cluster with 20 \msun~BHs for comparison (dotted lines).}
\label{fig:cv_tmerge}
\end{figure}

For a given hardness $x$ obtained by Eq.~\eqref{eq:hardness}, the semi-major axis of an ejected binary depends only on the velocity dispersion of a cluster. Therefore, we obtain a semi-major axis and merging time-scale of a binary as a function of $\sigma$. It is expected that ejected binaries from the cluster with larger velocity dispersion have shorter semi-major axes ($x\propto 1/a\sigma^{2}$). Thus, the merging time-scale of ejected binaries from a cluster with larger $\sigma$ tends to be shorter. The lower panel of Fig.~\ref{fig:cv_tmerge} illustrates this. For all ejected binaries, we calculate their merging time-scale using Eq.~\eqref{eq:tmerge}. The ratio of merging binaries to ejected BH-BH or NS-NS binaries ($R_{\rm M}$) increases as the central velocity of the GC increases. The results are not sensitive to the cluster concentration. The upper panel of Fig.~\ref{fig:cv_tmerge} shows a distribution of central velocity dispersions of MW GCs \citep{gne02}. \citet{gne02} calculated the central velocity dispersions using the King model assuming a constant mass-to-light ratio. The median central velocity of MW GCs is about 7.5 km s$^{-1}$. At this velocity, about 40 and 15 per cent of NS-NS and BH-BH ejected binaries are expected to merge within 13 Gyr.

The merger rate of BH-BH or NS-NS binaries per GC can be obtained as follows
\begin{equation}  
\label{eq:r_gc}
{\cal R}_{\rm GC}=\sum_{i=1}^{n} \frac{M_{\rm i}}{\langle m_{\rm *}\rangle} f_{\rm n} f_{\rm e,b} R_{\rm M}(\sigma_{i}(0)) /(n t_{\rm GC})~,
\end{equation}
where $M_{\rm i}$ is the mass of each GC which is given in the Gnedin's catalogue. Thus, $M_{\rm i}/\langle m_{\rm *}\rangle$ represents the number of stars in each GC. $f_{\rm n}$ is the number fraction of BHs or NSs in a GC. For clusters containing BHs, we obtain $f_{\rm n}\approx0.0018$ based on the Kroupa initial mass function \citep{kro01}. If the Scalo IMF \citep{sca86} is assumed, we have $f_{\rm n}\approx0.0007$. In this work, we assume $f_{\rm n}=0.001$ as a reference value for the BH number fraction. The number fraction of NSs is set to be $f_{\rm n}=0.005$, given by the Kroupa IMF. $f_{\rm e,b}$ is the number fraction of BH-BH or NS-NS binaries among ejected BHs or NSs. As we describe in \S \ref{sec:results}, about 30 per cent of ejected populations are in the form of BH-BH or NS-NS binaries. This implies $f_{\rm e,b}=0.3/2=0.15$. The factor of $1/2$ is applied, because an ejected binary is composed of two BHs or NSs. The merging rate $R_{\rm M}$ is written as a function of central velocity and is obtained from Fig.~\ref{fig:cv_tmerge}. A mean value of $R_{\rm M}$ obtained from models with $W_{\rm 0}=6$ and $9$ is used here. The total number of GCs in the catalogue is $n=141$. We assume that all GCs were formed 13 Gyr ago ($t_{\rm GC}=13$ Gyr) and the formation rate of compact binary mergers per GC is uniform.

The number fraction of ejected compact objects can affect our results. In this work, we assume that all BHs are ejected from a cluster. This assumption is plausible because it is thought that most BHs are ejected from GCs through dynamical interactions \citep*{kul93,sigh93,por00,aar12}. As presented in Fig.~\ref{fig:nesc_bb}, formation and evolution of BH-BH binaries begin at early evolution phases in a cluster's life, i.e., before a half-mass relaxation time. In addition, significant number of BHs are depleted in a short time ($5-10$ $t_{\rm rh}(0)$ for $f_{\rm BH}=2$ per cent models). Thus, it is likely that almost all BH-BH binaries have been ejected from GCs at the current epoch. The estimates of BH-BH mergers in Table \ref{tab:rates} are obtained using this assumption. However, recently, \citet{str12} detected two stellar mass BH candidates in M22 and several studies predicted that some fraction of BHs can remain in the clusters. The effects of remaining BHs in GCs on our estimates will be discussed in \S \ref{sec:discussions}.

Unlike BH cases, not all NSs are expected to be ejected from a cluster. This is because NS-NS binary ejection occurs much later, i.e., after almost all BHs are ejected (recall Fig.~\ref{fig:posi_nesc}). Therefore, the evolution phase of a cluster has to be taken into account when calculating $R_{\rm M}$. We assume all NSs are ejected from core-collapsed GCs. Among known GCs in MW, 29 GCs are considered to be core-collapsed \citep{djo86,har10} and we obtain $R_{\rm M}$ from Fig.~\ref{fig:cv_tmerge} for these clusters only. Their masses contribute less than 20 per cent of the total mass of GCs in \citet{gne02} catalogue. On the other hand, no NS-NS binary ejection is assumed for GCs that have not experienced core collapse. Therefore, their contributions to NS-NS binary merging rate are neglected ($R_{\rm M}=0$).

\begin{table*}
\caption{Estimated merger rates per GC for NS-NS and BH-BH binaries ejected from GCs. We present (a) the conservative rate estimates, (b) reference rate estimates based on our reference model, and (c) optimistic rate estimates. Within the parentheses, we also show the corresponding detection rates with the advanced LIGO-Virgo network. See text for details on calculations and assumptions.} 
\label{tab:rates}
\begin{tabular*}{0.9\textwidth}{@{\extracolsep{\fill}}cccc}
\toprule
        & conservative            & reference               & optimistic   \\ 
\midrule       
        & ${\cal R}_{\rm con}$ (${\cal R}_{\rm gw,con}$) &  ${\cal R}_{\rm re}$ (${\cal R}_{\rm gw,re}$)& ${\cal R}_{\rm opt}$ (${\cal R}_{\rm gw,opt}$)  \\
   & GC$^{-1}$ Gyr$^{-1}$ (yr$^{-1}$) &  GC$^{-1}$ Gyr$^{-1}$ (yr$^{-1}$) & GC$^{-1}$ Gyr$^{-1}$ (yr$^{-1}$)  \\
\midrule
BH-BH  &  n/a  & 2.5 (15) & 10 (60)   \\
\midrule
NS-NS  & 0.11 (0.01)  & 0.27 (0.024) & 1.1 (0.1)  \\
\bottomrule
\end{tabular*}
\end{table*}

Our cluster models do not involve natal kicks and all NSs and BHs are assumed to be retained in GCs after supernova explosions. However, if natal kicks are included in a cluster model, total number of NSs in a cluster is expected to be smaller than what is used in this work. \citet{dru96} obtained the retention fraction less than 4 per cent for single NSs and 2 to 5 times higher for binary systems through Fokker-Planck simulations. Our `reference' NS-NS merger rate estimate in Table \ref{tab:rates} is obtained by assuming a retention fraction of 10 per cent \citep*{pfa02}. The `conservative' values for NS-NS merger in Table \ref{tab:rates} are based on the 4 per cent suggested by \citet{dru96} for single NSs. On the other hand, assumptions on kicks do not significantly affect our results of BH-BH merger rate estimates. The kick velocity of a BH will be reduced by the mass ratio between a NS and a BH, if momentum given to a BH is same with that given to a NS. The estimated values of BH-BH merger in Table \ref{tab:rates} are obtained by assuming the full retention of BHs.

Using the above results, we calculate the detection rate of NS-NS and BH-BH mergers with the advanced LIGO-Virgo network. Following \citet*{ole06,bel07,ban10,dow11}, the detection rate of compact binary mergers can be obtained as follows
\begin{equation}
\label{eq:r_detect}
{\cal R}_{\rm det}= {\cal R}_{\rm GC} \rho_{\rm GC} 4\pi \int \frac{r(z)^2}{1+z} \frac{dr}{dz} dz ~,
\end{equation}
where ${\cal R}_{\rm GC}$ is the merging rate of compact binaries per GC, $\rho_{\rm GC}$ is the spatial density of GCs, and $r$ is a comoving distance. ${\cal R}_{\rm GC}$ is obtained by Eq.~\eqref{eq:r_gc}. We adopt $\rho_{\rm GC}=8.4 h^3$ Mpc$^{-3}$ from \citet{por00}, where the authors calculated the spatial density of GCs taking into account the spatial density as well as specific frequencies of GCs for various types of galaxies. The $(1+z)^{-1}$ term is introduced to account for the cosmological time dilation. By integrating Eq.~\eqref{eq:r_detect} up to the observable distance of a detector, we can obtain the detection rate of merging NS-NS and BH-BH ejected from GCs. \citet{aba10} presented that the horizon distances of advanced LIGO-Virgo network for NS-NS and BH-BH mergers to be $\rm D_{\rm H}\approx$445 Mpc and 2187 Mpc, respectively. Here, we use the volume averaged distance ($\rm D_{\rm V}$) as a limit of integration in Eq.~\eqref{eq:r_detect}. $\rm D_{\rm V}$ is obtained from the horizon distance ($\rm D_{\rm H}$) dividing by correction factor ($\approx2.26$), which is obtained by averaging over all orientations and sky locations. Based on our reference merger rate estimates, the detection rates of advanced LIGO-Virgo network are obtained to be $\sim0.024$ yr$^{-1}$ (NS-NS) and $\sim15$ yr$^{-1}$ (BH-BH), respectively. The cosmological parameters $\Omega_{m}=0.27$, $\Omega_{\Lambda}=0.73$ and $h=0.72$ are used in this work. We note that Planck 2013 results \citep{pla13} presented a smaller Hubble constant as well as $\Omega_{\Lambda}$ and a larger $\Omega_{m}$ than the parameters used in this work. This implies lower detection rates by $10-15$ per cent from the values presented in Table \ref{tab:rates}.

Our rate estimations are based on conservative assumptions such as the average mass of stars in a cluster. The lightest stellar populations are fainter, and their fraction is not well constrained. In this work, the average mass of stars in GC is about \avgm=0.7 \msun. However, \avgm=0.56 \msun~can be obtained by \citet{sca86} and \avgm=0.33 \msun~was used in \citet{djo93}. The merger rate is inversely proportional to \avgm~(see Eq.~\eqref{eq:r_gc}), and therefore, our rate estimates can be increased by $25-100$ per cent, if assuming smaller \avgm~used in \citet{sca86} and \citet{djo93}.

Our work consider the present-day GCs, which are presumably only a fraction of the original population. Some of the GCs could have been completely destroyed in early phase of host galaxy or have lost a significant number of stars by Galactic tidal field or tidal shock \citep{lee95,gne97}. Although the IMF of GCs is highly uncertain, our results are likely to be underestimated by neglecting the contribution from these `evaporated' clusters. It is expected that original population of GCs is several times of that of present GCs \citep*{bau98,ves98,shi08,shi13}. Notably, it is known that about 50 per cent of GCs can be dissolve in the current age of the universe by the tidal field of the host galaxy \citep{lee95,tak00}. If the mass loss process has been continuous over the age of the host galaxy, we can assume that the present GCs are only the half of the original population. If this is the case, the number of clusters that have produced merging NS-NS and BH-BH binaries are about twice more than what we consider in this work. Based on what is discussed, we calculate our `optimistic' rate estimates by multiplying a factor four to our reference value. 

Note that we have ignored the effects of initial rotation, which could significantly accelerate the course of dynamical evolution (see, \citet{hong13} for recent study on rotating clusters, and references therein). The current population of GCs does not show dynamically significant amount of rotation, but that does not mean that the clusters were born without rotation. If initial rotation was important, probably more fraction of GCs could have been completely destroyed by now. Although we do not contain this factor when we estimate merger rates, actual number of binaries originated from GCs could be more than our estimates.

The `conservative', `reference' and `optimistic' merger rates and corresponding detection rates for BH-BH and NS-NS mergers are listed in Table~\ref{tab:rates}. Including the natal kicks, we suggest reference rate estimates of BH-BH and NS-NS binaries per GC as 2.5 and 0.27 GC$^{-1}$ Gyr$^{-1}$, respectively. Considering that there are about 200 GCs in MW, 0.5 BH-BH merging and 0.054 NS-NS merging per Myr are expected in MW.

\section{Discussions}\label{sec:discussions}

In this work, we consider the effects of BHs and NSs on dynamical formation of binaries in GCs by performing N-body simulations. We estimate GW detection rates relevant to BH-BH and NS-NS mergers that are ejected from clusters with the advanced GW detector network. Followings are main differences between our work and previous studies. Firstly, we consider the effects of NSs on dynamical formation of binaries in GCs. We estimate the detection rates of mergers containing NSs quantitatively. Secondly, we calculate the fraction of compact binary mergers among the ejected as a function of the velocity dispersion of clusters. Previous studies often calculated merger rates from the cluster models with fixed central densities. Thirdly, we consider the effects of primordial binaries in N-body simulation, which were not included in \citet{ban10} and \citet{tan13}. Lastly, we examine the King models with different concentrations ($W_{0}=6$ and 9), not only changing physical scales at a given concentration.

Despite these differences in method and assumptions, the estimated GW detection rates for BH-BH mergers from this work are consistent with previous studies. \citet{ban10} used \textsc{nbody}6 code for an isolated Plummer model, varying the BH fractions in a cluster and assumed a reflective boundary for some models in order for mimicking the core of a cluster. They also considered 10 \msun~BHs and changed the number fraction of BHs. They obtained the detection rate of advanced LIGO for BH-BH binaries to be $31\pm7$ yr$^{-1}$ and concluded that the intermediate-age massive clusters are likely to produce observable BH-BH mergers. \citet{dow11} obtained the detection rates up to 15 and 29 yr$^{-1}$ for advanced detectors using eccentricities from a Monte Carlo code and thermal distribution, respectively. They considered stellar evolutions for singles and binaries. They have examined 16 cluster models which have different primordial binary fractions, metallicities and initial concentrations. \citet{tan13} obtained the detection rate of BH-BH mergers from escapers as $0.5-20$ yr$^{-1}$ for 50 per cent retention fraction. He used \textsc{nbody}4 code and considered a BH mass function peaked at around 5 \msun. This is obtained from a stellar evolution model \citep*{hur00, hur01, eld04} and assumption of retention fraction which is independent of zero-age main-sequence stars and remnant masses. Considering 8k$-$128k stars, he investigated cluster models with different initial densities inside half mass radii and BH retention fractions. Then, he extrapolated the results to larger ($N\sim10^{6}$) clusters and estimated the detection rate of BH-BH mergers ejected from clusters.

Our detection rates are also roughly consistent with the results obtained by \citet{ole06} and \citet{sad08}. \citet{ole06} predicted a few BH-BH binary merger detections per yr by advanced LIGO, assuming that BHs are concentrated in the core of a cluster and they are completely decoupled from the other stars. In addition to studies on dynamically formed binary mergers, \citet{sad08} performed population synthesis approach. Considering both field and cluster populations, they obtained GW detection rates of BH-BH binary mergers between 25 and 3000 per yr with the advanced LIGO. They found that the detection rate is dominated by BH-BH mergers originated from clusters when initial mass fraction contained in clusters is larger than 0.001. They assumed that BHs in clusters remain in thermal equilibrium with other stars.

All studies mentioned above, including our work, agree that the contribution from BH-BH mergers ejected from MW GCs is comparable to or can be exceeding those formed in the disk. Applying population synthesis method, \citet{bel07} calculated the merger rate of BH-BH binaries formed from a MW-like galaxy. They suggested Galactic BH-BH merger rate, based on a disk population, is $0.02-0.03$ Myr$^{-1}$. The corresponding detection rate with advanced LIGO was expected to be $1.6-2.5$ yr$^{-1}$. More recently, the review paper by the LIGO-Virgo collaboration \citep{aba10} presented the detection rate for BH-BH binaries with the advanced LIGO-Virgo network to be 20 yr$^{-1}$. On the other hand, recent work by \citet{dominik2012} suggested much larger BH-BH merger rate estimates than other groups. Based on the latest version of the code used in \citet{bel07}, they performed population synthesis of the BH-BH binaries in the field and obtained Galactic BH-BH merger rate to be 1.9 or 8.2 per Myr for Solar metallicity environment for two different models. For models with sub-Solar metallicity, they expected BH-BH merger rate, formed in the field, to be 13.6 or 73.3 per Myr.

Considering observed properties of the only known NS-NS binary (PSR J2127-11C) discovered in M15 \citep{prince91}, \citet{phinney91} estimated the merger rate per volume of cluster-origin NS-NS binaris to be $3\times 10^{-9} h^{3}$ Mpc$^{-3}$ yr$^{-1}$. Based on his result, it is generally assumed that the Galactic NS-NS merger rate is dominated by the field population and that from GCs is negligible. We confirm this is indeed the case. The estimated detection rate of NS-NS binary mergers ejected from GCs contributes less than 0.1 per cent of that from field mergers ($\sim40$ yr$^{-1}$ by \citealt{aba10}).

In \S \ref{sec:rates}, we assume that all BHs have been ejected from GCs when we calculate the merger rate of BH-BH or NS-NS binaries. However, some studies predicted that BHs can be retained in clusters at the current epoch. \citet{mac08} performed N-body simulations with $N\sim10^{5}$. They chose initial parameters that are representative for youngest massive star clusters found in the Magellanic Clouds. They found up to 50 per cent of BHs can be retained in a cluster after 10 Gyr, corresponding to a primordially mass-segregated model with no natal kicks (Run 4 in \citealt{mac08}).

The discovery of two stellar-mass BH candidates ($10-20$~\msun) in radio continuum images of M22 is the first observational evidence found in MW GCs \citep{str12}. Considering that $2-40$ per cent of BHs in a cluster are belonged to the binaries with observable mass-accretion over 10 Gyr \citep{iva10}, \citet{str12} argued that there are $\sim5-100$ BH binaries in M22 inferred from the two detections of BHs. Qualitatively, the existence of BHs in M22 provide good explanation for its core radius which is larger than other MW GCs (e.g., \citealt{mac08}). Motivated by the discovery and claims made by \citet{str12}, \citet{sip13} performed large-scale N-body simulations using a cluster model which is consistent with M22. They consider a cluster with $N=262 500$ stars (binary fraction of 0.05, metallicity of 0.001), applying $\sim 10$ per cent of retention rate for BH and NS after natal kicks. Their model involves stellar evolution and natal kicks. They found that about 1/3 of BHs remain in the core after 12 Gyr.

\citet{mor13} also suggested that large fraction of BHs can be retained in clusters. They performed Monte Carlo simulations with $N=3\times10^{5}$ stars following the King model with $W_{0}=5$. They chose the metallicity of Z=0.001 and primordial binary fraction as $f_{\rm b}=0.1$. Their model contains about 700 BHs initially, following a BH mass spectrum obtained from \citet{bel02} in a range between $\sim5-30$~\msun. They found that the most massive BHs ($\ge20$~\msun) are ejected preferentially due to their higher interaction rates, similar to the BHs and NSs in this work (Fig.~\ref{fig:posi_nesc}), and more than half of BHs are  retained in a GC after 12 Gyr.

These works demonstrated that $\sim30-50$ per cent BHs can be retained in massive GCs, which have large half-mass relaxation time \citep{mac08,sip13} or low concentration \citep{mor13} with BH mass distribution. Among those listed in the Harris' and Gnedin's catalogues, more than 30 GCs have half-mass relaxation time greater than 2 Gyr. These cluster contribute about 40 per cent of the total cluster mass in MW. In order to estimate the effect of remaining BHs, we calculate the BH-BH merger rate by multiplying the fraction of ejected BHs from GCs to Eq.~\eqref{eq:r_gc}. The rate estimate is reduced by 20 per cent if we assume that only the half of BHs are ejected from GCs which have the half-mass relaxation time longer than 2 Gyr and all BHs are ejected from the other GCs. Furthermore, if we accept \citet{cha13} which claims longer half-mass relaxation time of GCs than Harris' catalogue, and assume that half of BHs remain in non core-collapsed GCs, our rate estimate is reduced by about 40 per cent.

In addition to remaining BHs, mass spectrum also affects to cluster dynamics and the fate of the ejected binaries. Our assumption of NS mass (1.4 \msun) is plausible because it is known that NSs have very narrow mass distribution \citep*{kiz10, sch10, val11, zha11, ozel12}. On the other hand, BHs have relatively broader range of masses. Analysing 16 BHs in transient low-mass X-ray binaries in MW, \citet{ozel10} found that the mass distribution of BHs in MW can be described by Gaussian function with a mean of 7.8 \msun~and a standard deviation of 1.2 \msun. \citet{farr11} obtained similar results for low-mass X-ray binaries, but they also considered more massive BHs containing high-mass X-ray binaries. In addition, \citet{bel10} showed that the BHs whose masses are up to 80 \msun~can be formed in the low metallicity environment such as GC. Different BH mass spectra are assumed in cluster modellings. For example, \citet{mor13} used the BH masses with two peak around 10 and 20 \msun~ according to \citet*{bel02} for metallicity $Z=0.001$. While \citet{tan13} used a single peaked mass spectrum as described earlier. A BH mass spectrum in clusters implies a various range of detectable distances for BH-BH mergers by GW detection.

In order to examine the effects of BH mass spectrum, we performed N-body simulations with 20 \msun~BHs in a cluster assuming two concentrations ($W_{0}=6$ and 9). We find that the mean hardness of 20 \msun~BH-BH binaries ejected from a cluster is about two times larger than that of binaries with 10 \msun~BHs. Then, their semi-major axes and merging times can be obtained by Eq.~\eqref{eq:hardness} and Eq.~\eqref{eq:tmerge} and their merging rates against central velocity dispersion of GC are presented in Fig.~\ref{fig:cv_tmerge}. As shown in Fig.~\ref{fig:cv_tmerge}, the merging rates of ejected 20 \msun~BH binaries are very similar to that of 10 \msun~BH binaries, though they are slightly smaller. The gap of merging rates between $W_{0}=6$ and 9 becomes narrower because the increase rate of hardnesses of $W_{0}=9$ is a little bit larger than that of $W_{0}=6$.

Therefore, the detection rate of 20 \msun~BH-BH mergers are largely determined by the detection range of GW detectors. Taking into account their GW frequencies and sensitivities of advanced detectors (e.g., \citealt{cut94,ole06}), the detectable distance of 20 \msun~BH merger is roughly twice larger than that of 10 \msun~BH merger. If we fix the BH mass to be 20 \msun~and use the merging rate presented in Fig.~\ref{fig:cv_tmerge} and GW detection range of advanced detectors for 20 \msun~BHs, the GW detection rate for BH-BH mergers increases about a factor three. In reality, a cluster contains both 10 \msun~and 20 \msun~BHs. If we assume the number ratio between them to be $2:1$, which is roughly consistent with Figure 1 in \citet{mor13}, about 25 BH mergers are expected to be detected by advanced detector networks for reference value. This implies that our estimation of detection rate can be increased by a factor of two when considering the BH mass distribution.

Based on our modelling, we find that no BH-NS binaris are formed from a cluster. This is because almost all BHs are depleted when NSs start to sink toward the centre. If we allow a BH mass spectrum, however, BH-NS binaries  can be formed in a GC. The formation rate of BH-NS binaries is sensitive to model assumptions. Within the uncertainties in cluster modelling and BH population in clusters, we remark that BH-NS mergers with cluster origin are likely to be rare, but not completely ruled out \citep*{cla13}.

Lastly, we remark on the implication of our results in the context of the gamma-ray burst modelling. The offsets of short-duration gamma-ray bursts (GRBs) relative to their host galaxies are known to be larger than those of long-duration GRBs. The median value of projected offsets distribution of short GRBs is $\sim$5 kpc, which is about 5 times larger than that of long GRBs \citep*{fon10}. Some short GRBs have offsets greater than 20 kpc, even for GRB 060502B has projected offsets of $73\pm19$ kpc \citep{blo07}. These large offsets of short GRBs can be explained by the escape velocity of compact binaries formed in GCs. Some of the ejected binaries have velocities over $150\sim200$ km s$^{-1}$ as we see in \S \ref{sec:results}. These fast-moving binaries, if mergers, are expected to move out of several tens of kpc from host galaxy if they are ejected at early times. Some of the short GRBs with large offsets may be explained by these.

\section*{Acknowledgements}
This work was supported by the National Research Foundation of Korea (NRF) grant NRF-2006-0093852 funded by the Korean government. HML acknowledges Alexander von Humboldt Foundation for the Research Award in 2013. We also thank Sambaran Banerjee for careful reading of the manuscript.


\label{lastpage}

\end{document}